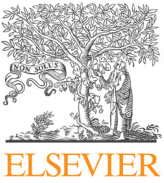
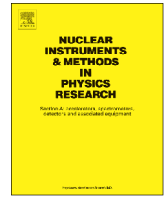
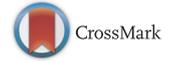

# In-situ study of light production and transport in phonon/light detector modules for dark matter search

M. Kiefer [a,*], G. Angloher [a], A. Bento [b], C. Bucci [c], L. Canonica [c], A. Erb [d,e], F.v. Feilitzsch [d], N. Ferreiro Iachellini [a], P. Gorla [c], A. Gütlein [f,g], D. Hauff [a], J. Jochum [h], H. Kluck [f,g], H. Kraus [i], J.-C. Lanfranchi [d], J. Loebell [h], A. Münster [d], F. Petricca [a], W. Potzel [d], F. Pröbst [a], F. Reindl [a], S. Roth [d], K. Rottler [h], C. Sailer [h], K. Schäffner [c,j], J. Schieck [f,g], S. Schönert [d], W. Seidel [a], M.v. Sivers [d], L. Stodolsky [a], C. Strandhagen [h], R. Strauss [a], A. Tanzke [a], C. Türkoğlu [f,g], M. Uffinger [h], A. Ulrich [d], I. Usherov [h], S. Wawoczny [d], M. Willers [d], M. Wüstrich [a], A. Zöller [d]

[a] Max-Planck-Institut für Physik, D-80805 München, Germany
[b] CIUC, Departamento de Fisica, Universidade de Coimbra, P3004 516 Coimbra, Portugal
[c] INFN, Laboratori Nazionali del Gran Sasso, I-67010 Assergi, Italy
[d] Physik-Department, Technische Universität München, D-85748 Garching, Germany
[e] Walther-Meißner-Institut für Tieftemperaturforschung, D-85748 Garching, Germany
[f] Institut für Hochenergiephysik der Österreichischen Akademie der Wissenschaften, A-1050 Wien, Austria
[g] Atominstitut, Vienna University of Technology, A-1020 Wien, Austria
[h] Physikalisches Institut, Eberhard-Karls-Universität Tübingen, D-72076 Tübingen, Germany
[i] Department of Physics, University of Oxford, Oxford OX1 3RH, United Kingdom
[j] Gran Sasso Science Institute, I-67100 L'Aquila, Italy

## ABSTRACT

The CRESST experiment (Cryogenic Rare Event Search with Superconducting Thermometers) searches for dark matter via the phonon and light signals of elastic scattering processes in scintillating crystals. The discrimination between a possible dark matter signal and background is based on the light yield.

We present a new method for evaluating the two characteristics of a phonon/light detector module that determine how much of the deposited energy is converted to scintillation light and how efficiently a module detects the produced light. In contrast to former approaches with dedicated setups, we developed a method which allows us to use data taken with the cryogenic setup, during a dark matter search phase. In this way, we accounted for the entire process that occurs in a detector module, and obtained information on the light emission of the crystal as well as information on the performance of the module (light transport and detection).

We found that with the detectors operated in CRESST-II phase 1, about 20% of the produced scintillation light is detected. A part of the light is likely absorbed by creating meta-stable excitations in the scintillating crystals. The light not detected is not absorbed entirely, as an additional light detector can help to increase the fraction of detected light.

## 1. Introduction

### 1.1. General context

The CRESST (Cryogenic Rare Event Search with Superconducting Thermometers) dark matter experiment aims at detecting WIMP-nucleus scattering [1] in inorganic scintillating crystals operated as cryogenic detectors. Energy deposited in the crystals creates phonons and scintillation light.

The phonon signal is measured by a transition-edge sensor (TES) evaporated onto the scintillating crystal. The scintillation light is detected by a separate light detector also read out by a TES [2]. The TESs in CRESST consist of a tungsten thin-film structure thermally stabilized at the transition between normal and superconducting state. In this regime, even the very small temperature variations $\mathcal{O}(\mu K)$ caused by individual particles depositing energy

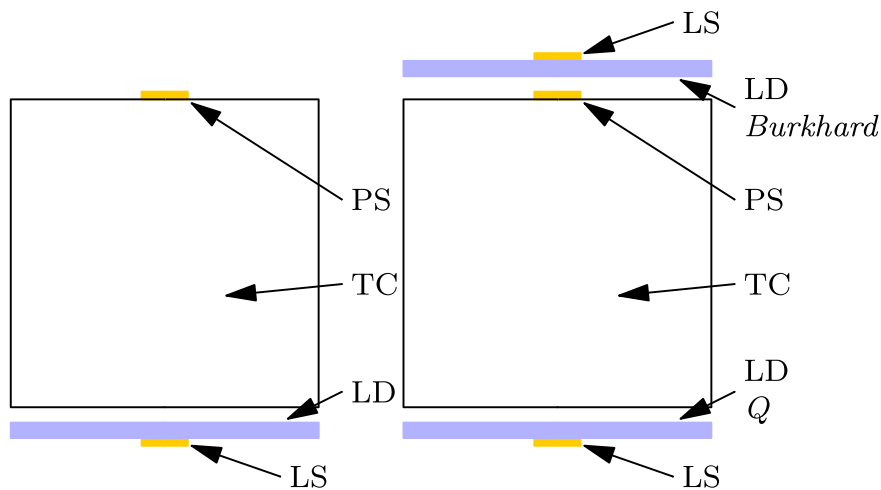

**Fig. 1.** Schematic drawing of the detector module designs considered in this work. In all modules, the interaction in the target crystal (TC) produces phonons and scintillation light. The phonons are read out by the phonon TES (PS). The light is absorbed in the light detector (LD) and read out by the light TES (LS). Target crystal and light detector are surrounded by a scintillating reflective foil (not shown here). The left sketch shows a detector of the standard design. The double light detector module (right) was equipped with two light detectors, LD Q and LD Burkhard. This non-standard design serves to investigate how scintillation light propagates within a module.

in the detector result in resistance changes that we can measure with a SQUID amplifier.

The scintillating crystal and the light detector are surrounded by a housing made of a scintillating and reflective foil, which (apart from masking certain types of background events by emitting light, cf. [3]) prevents light from escaping without contributing to the signal. The ensemble of crystal, light detector, foil and a surrounding copper structure is called a detector module.

The signal of the phonon channel is a measure of the deposited energy. The fraction of deposited energy that is converted to scintillation light depends on the nature of the interacting particle (e.g. αs or γs). Hence, the ratio of scintillation light to deposited energy is used to distinguish between different types of interacting particles. Increasing the amount of detected scintillation light per deposited energy is crucial for maximizing the background suppression capabilities of the detector modules —and hence the overall sensitivity of the experiment.

Only a fraction of the produced light is actually detected. This is due to the limited transparency of the crystal (which only partially emits the produced light), due to the geometry and the efficiency of the reflector as well as due to the size and absorptivity of the absorber of the light detector. Therefore, disentangling the factors that affect production and detection of scintillation light helps to identify the key factors for further improving the light signal and hence signal-background discrimination of experiments using the phonon-light technique to look for dark matter as well as for neutrinoless double-beta decay [4,5] or [6].

### 1.2. Concept of the new evaluation method

The typical, established methods for characterizing the crystals work by irradiating the crystal in a dedicated setup at room temperature and by measuring the scintillation light, e.g. with a photomultiplier tube. With the new method (for a detailed description cf. Section 2.3), we use data acquired in situ, with the cryogenic setup during dark matter data taking to determine the efficiencies at which light is being produced and detected.

We use two clearly identified lines originating from interacting particles of different nature (i.e. one α-line and one γ-line). Then, we know the energy deposited in the crystal (the Q-values of the lines), and we can measure the absolute energy deposited in the light detector. The energy of the phonon detectors is not on an absolute scale as it depends (and thus contains information) on how much of the deposited energy is converted into scintillation light. The fraction of light that is lost is the same in both cases. With these information, we derive production and detection characteristics of the detectors by evaluating the Q-values and phonon and light detector readings of the two lines.

## 2. Experimental setup, data and analysis

To verify the findings and in order to obtain more detailed information on how to possibly optimize the detector modules, we evaluated light production and light detection of two different detector designs. This section describes the designs we investigated and the method by which we analyzed the data.

### 2.1. Detector designs

All data analyzed here have been acquired in CRESST-II phase 1 [3]. The detector modules of the standard design (cf. Fig. 1, left) consist of a target crystal (in which the energy is deposited) equipped with a TES that detects the phonons. The crystals are usually calcium tungstate cylinders of 40 mm height and diameter. The crystal face opposite to the TES is roughened in order to facilitate light propagation towards the light detector.

The double light detector module depicted on the right in Fig. 1 allows an additional insight in the light propagation within the module. The light detector called Q is located adjacent to the roughened surface of the scintillating crystal. The light detector named Burkhard faces the crystal at the opposite side, near the phonon TES.

If the two light detectors in sum detected more light than a single detector,[1] this would suggest that the situation could possibly be improved by changing the light detectors and/or the path of the light within a module. If introducing the second light detector did not significantly rise the total amount of detected light (so that already a single light detector gathers all the available light), this would indicate that the reflectivity of the housing, the absorptivity of the light detector and the transparency of the crystal were nearly optimal.

---

[1] With the fact taken into account that individual crystals differ in light production efficiency.

## 2.2. Calibration sources

Energy depositions in the phonon and the light detectors result in voltage pulses delivered by the read-out system [7]. The height of the pulses is a measure of the deposited energy, which is calibrated using radioactive sources:

- We calibrated the phonon detector against the α-decay of $^{147}$Sm at 2310.5 keV[2] [8]. Linearity of the energy scale to lower energies was confirmed by using an external $^{232}$Th-source [9] emitting γ-lines at 338, 583, 911, 1588, and 2614 keV.
- The light detectors are directly exposed to low-energy X-rays from an external $^{55}$Fe-source, setting the absolute energy scale. In this case, we used electrical pulses injected into heaters connected to the TESs to confirm linearity of the detector response. The calibration source emits single photons of 5.9 and 6.5 keV, while scintillation light of the same energy consists of many photons in the eV-range. The signals of the calibration source and of the scintillation events have a slightly different pulse shape due to the long scintillation time $\mathcal{O}(360\ \mu s)$ [10] of the crystals in use. We took this difference into account via the ratio between pulse height and integral of the pulse for the different event classes.

## 2.3. Analysis method

We consider the energy flow in the detector module as illustrated in Fig. 2: The fraction of deposited energy that is converted into scintillation light depends on the nature of the particle and the quality of the crystal. The fraction of the produced light that is finally detected depends on the transparency of the crystal, on the geometry and efficiency of the reflector as well as on size and absorptivity of the absorber of the light detector.

At the beginning of the process, a particle deposits an energy $Q_{dep}$ in the scintillating crystal. Depending on the particle type (indicated by $\phi$), this energy divides up into a relatively small fraction leaving the crystal as scintillation light $Q_S = \mathcal{F}_\phi Q_{dep}$ and into a part remaining in the crystal as phonons $Q_P = (1 - \mathcal{F}_\phi) Q_{dep}$. The term $\mathcal{F}_\phi$ denotes the fraction of energy deposited by particle $\phi$ converted to light. It depends on the particle type and on the characteristics of the crystal and is defined as

$$\mathcal{F}_\phi = \mathcal{R}_\phi \epsilon_s. \tag{1}$$

The term $\epsilon_s$ describes the scintillation efficiency of the crystal, namely the fraction of energy that a crystal converts to scintillation light when it is excited by a γ-quantum. The particle-dependent part can be described [11] by a relative light yield[3] $\mathcal{R}_\phi$ which is the light yield of $\phi$-particles relative to the light yield of γ-quanta. As the light yield of any particle $\xi$ is defined as the light energy over the deposited energy $E_{L,\xi}/Q_{dep,\xi}$, the relative light yield is given by

$$\mathcal{R}_\phi = \frac{E_{L,\phi}/Q_{dep,\phi}}{E_{L,\gamma}/Q_{dep,\gamma}}. \tag{2}$$

We neglect effects like e.g. energy dependences, scintillator non-proportionality [13–15] or different behavior of electrons and γ-events [16] as they are only relevant at energies up to 100 keV, nearly one order of magnitude below the ones considered here.

Thus, the amount of energy converted into phonons[4] can be written as

$$Q_P = (1 - \mathcal{R}_\phi \epsilon_s) \cdot Q_{dep}. \tag{3}$$

The energy in the light channel divides up further. The detection efficiency $\epsilon_d$ describes how much of the light escaping the crystal is absorbed by the light detector.[5] The fraction $(1 - \epsilon_d)$ is lost, e.g. because the housing is not a perfect reflector.

The energy finally absorbed in the light detector can then be written as

$$Q_L = \mathcal{R}_\phi \epsilon_s \epsilon_d \cdot Q_{dep}. \tag{4}$$

If the phonon energies $Q_P$ were exactly known, Eq. (3) could be used to directly determine the scintillation efficiency $\epsilon_s$ of the crystals from a γ-measurement. However, we have to account for the fact that we calibrate the detectors with particles that create scintillation light. In general, a calibration is done by exposing the detector to particles of a well-known energy and setting the detector reading to that energy. In the phonon detector calibration, the energy that escapes as scintillation light is neglected as the phonon detector reading is set to indicate the entire amount of deposited energy. The detector reading and the phonon energy $Q_P$ are therefore off by a factor depending on the particle used for calibration.

To reconstruct physical energy depositions in the detectors $Q_{P,L}$ from the calibrated energy readings $E_{P,L}$, we model this calibration effect with the factors $\mathcal{C}_{P,L}$, for the phonon and the light channel respectively. The physical energy depositions $Q_{P,L}$ in the crystal thus result in energy readings $E_P$, $E_L$:

$$E_P = \mathcal{C}_P Q_P = \mathcal{C}_P (1 - \mathcal{R}_\phi \epsilon_s) \cdot Q_{dep} \tag{5}$$

$$E_L = \mathcal{C}_L Q_L = \mathcal{C}_L \mathcal{R}_\phi \epsilon_s \epsilon_d \cdot Q_{dep}. \tag{6}$$

As the energy of the aforementioned $^{55}$Fe source is directly absorbed in the light detector, we assume a calibration factor of $\mathcal{C}_L = 1$ for the light channel. In case of the phonon channel, we have to consider the energy that escapes as scintillation light. Since the phonon channel is calibrated using α-particles, the calibration factor for the phonon detector must be such that it compensates for the energy escaping as light in case of an α-deposition. This means that

$$\mathcal{C}_P = \frac{1}{1 - \mathcal{R}_\alpha \epsilon_s}. \tag{7}$$

Resolving Eqs. (5) (with $\mathcal{R}_\gamma = 1$, per definition as in Eq. (2)) and (7), the scintillation efficiency $\epsilon_s$ can be extracted from the phonon detector reading of a γ-line that was calibrated with an α-source:

$$\epsilon_s = \frac{\dfrac{1}{E_{P,\gamma}} - \dfrac{1}{Q_{dep,\gamma}}}{\dfrac{1}{E_{P,\gamma}} - \dfrac{1}{Q_{dep,\alpha}} \dfrac{E_{L,\alpha}}{E_{L,\gamma}}}. \tag{8}$$

The detection efficiency $\epsilon_d$ can be then determined from the light detector reading of the γ-event:

$$\epsilon_d = \frac{E_{L,\gamma}}{Q_{dep,\gamma} \epsilon_s}. \tag{9}$$

In summary, this method determines $\epsilon_s$ and $\epsilon_d$ by measuring the light signals ($E_{L,\gamma|\alpha}$) of a γ- and an α-event of known Q-values, as well as the phonon signal of the γ-event ($E_{P,\gamma}$). The Q-values of the α- and γ-decays have to be known.

---

[2] Since Sm is an intrinsic contamination, the energy deposition in the crystal corresponds to the Q-value of the decay.
[3] The concept is similar to the quenching factor [12] that considers two particles of the same $Q_{dep}$.
[4] We neglect energy lost in the phonon channel. If a process systematically pulled off a fraction of energy from the phonon system, the phonon detector calibration would cancel its effect.
[5] It is also possible that the crystal re-absorbs part of the propagating light and converts it back into phonons. As these are seen in the phonon channel, we consider this fraction of energy never having been emitted as light.

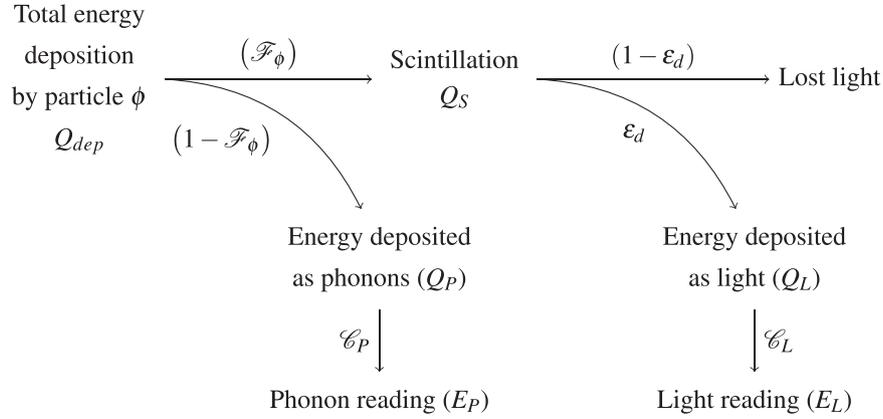

**Fig. 2.** Partitioning of the deposited energy in the detector module (see text for details). The deposited energy $Q_{dep}$ is shared between phonons ($Q_P$) and scintillation light ($Q_L$). The interpretation of the sensor readings $E_{P,L}$ depend on the calibration factors $\mathcal{C}_{P,L}$, taking into account the fraction $\mathcal{F}_\phi$ and the light detection efficiency $\epsilon_d$.

### 2.4. Data used for the analysis

To study the scintillation light signals of γ-events, we measured the two γ-lines of $^{228}$Ac and $^{212}$Bi at 726 and 727 keV. They originate from the $^{232}$Th source used for confirming the linearity of the detector response.

We used $^{180}$W ($Q = 2516$ keV [17]) to determine the light signals of α-particles. The crystals in use are tungstate crystals, hence this isotope is distributed uniformly through the volume,[6] like the calibration-α-source.

## 3. Results and discussion

We found that the light output of the modules is more strongly influenced by how well the crystals transport the scintillation light rather than how much of the energy they convert to light.

The values of $\epsilon_s$ indicate that, depending on the module, between 7.4% and 9.2% of the deposited energy is emitted as scintillation light (cf. Fig. 3). This agrees with the model featured in [10]. In case of single light detectors, a fraction between $\epsilon_d = 18\%$ and 28% of the produced light is detected (cf. Fig. 4).[7]

The values above the dashed lines are from the double light detector module, with its individual light detectors treated separately as well as in combination. When considering Q and *Burkhard* individually, they are the two light detectors with the lowest detection efficiencies. When one adds up their energies, the module Q/Verena/Burkhard has the highest detection efficiency $\epsilon_d$ of nearly 34%. The average for single light detector modules is ≈ 23%.

### 3.1. Consistency of the method

#### 3.1.1. Self-consistency

The double light detector module provides a check for the self-consistency of the method: the readings of *Burkhard* and Q are different (e.g. for the 727 keV-γ-line: $E_L = 9.9$ keV and $E_L = 8.4$ keV, respectively), which can be attributed to one of the light detectors facing the roughened and the other the polished surface of the crystal. Nonetheless, the method yields the same value for the scintillation efficiency $\epsilon_s$ for both light detectors while the detection efficiency $\epsilon_d$ differs. This confirms that characteristics related to the production of light can be distinguished from other effects in the detector modules.

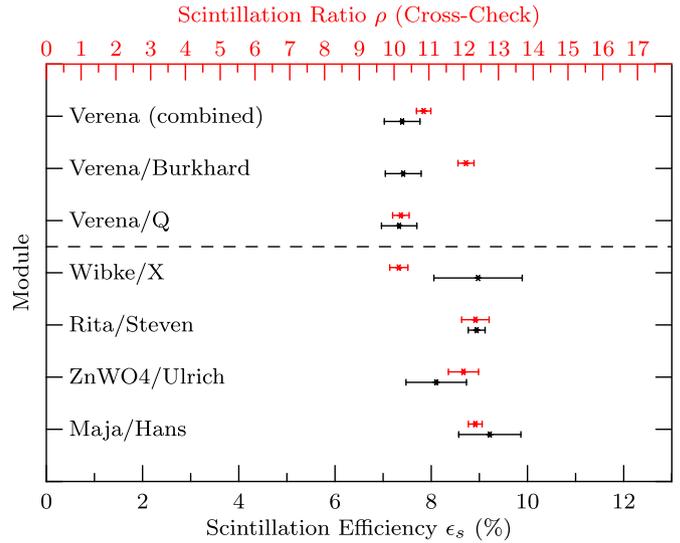

**Fig. 3.** Scintillation efficiencies obtained with the method introduced in this work (black). The dashed line separates the modules with a single light detector (below) from the module with a double light detector setup (above). Drawn in red are the axis and the data points obtained with the cross-check method explained in Section 4. (For interpretation of the references to color in this figure caption, the reader is referred to the web version of this paper.)

#### 3.1.2. Comparison with PMT measurements of the crystal scintillation

A simple approach to compare the relative amount of scintillation light of different crystals consists in irradiating the crystals at room temperature with a source of known energy and measuring the scintillation with a photo-multiplier tube (PMT), having a reference crystal to compare the results.

For all modules with a single light detector,[8] the values of the PMT measurements (red horizontal axis and data points in Fig. 5) and the products of scintillation and detection efficiency —the fraction of the deposited energy that is detected as light—(black horizontal axis and data points) are linearly correlated. Hence, temperature-dependent effects on how efficiently a crystal scintillates must be similar for all detectors, while the individual housings cannot introduce a large variation in terms of how well light is transported inside a module. From this, we conclude that (although being performed at room temperature) a PMT measurement is a good indicator for the crystal quality (i.e. the efficiencies at which light is produced and measured), and that all module housings investigated are on a similar level in terms of how they influence the light detection.

---

[6] By contrast, α-emitters associated with surface impurities would only probe a thin skin region due to the short range of α-particles.

[7] The systematic errors for the values are derived from uncertainties of the phonon and light detector calibration, the statistical errors from a fit of Gaussians to the peaks in the phonon and light energy spectrum, respectively. For the derived values, the errors are estimated using Gauss' law of error propagation. As the measured values within error are far away from the physical boundary of zero, we consider the errors as symmetrical.

[8] Inside the double light detector module, the light collection area is twice as large as in a module with a single light detector. Therefore, one cannot expect the

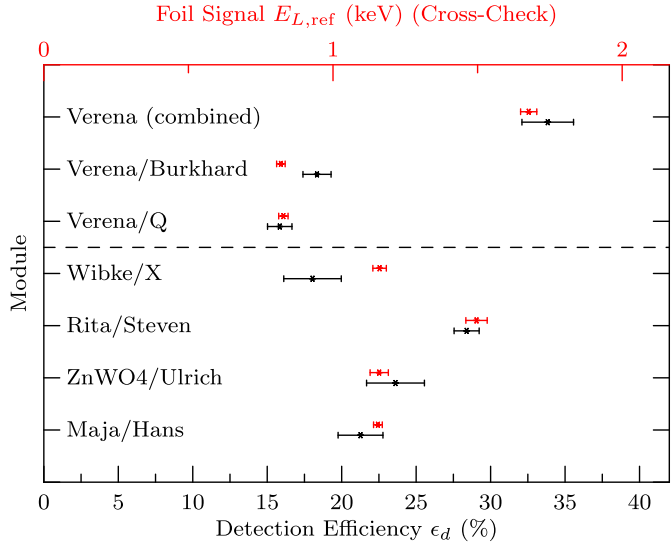

**Fig. 4.** Detection efficiencies obtained with the method introduced in this work (black). The dashed line separates the modules with a single light detector (below) from the module with a double light detector setup (above). Drawn in red are the axis and the data points obtained with the cross-check method explained in Section 4. (For interpretation of the references to color in this figure caption, the reader is referred to the web version of this paper.)

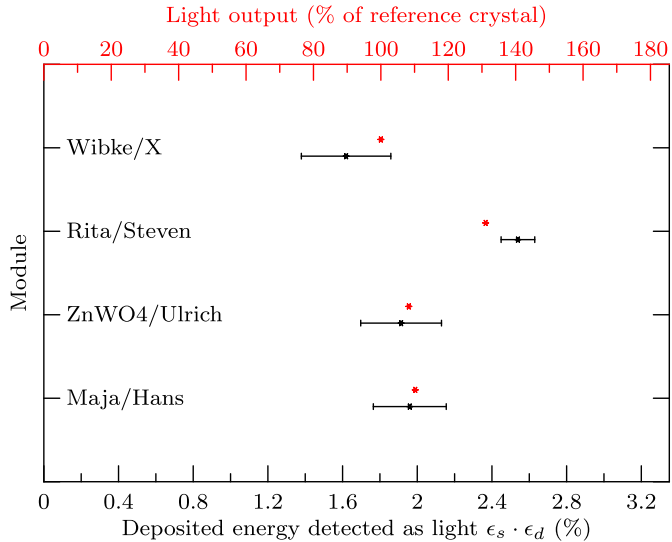

**Fig. 5.** The fraction ($\epsilon_s \cdot \epsilon_d$) of the deposited energy seen by the light detectors, obtained with the newly introduced method (black) and with the PMT method (red). (For interpretation of the references to color in this figure caption, the reader is referred to the web version of this paper.)

---

(*footnote continued*)
PMT-value and the product of scintillation and detection efficiency to have the same proportionality as in case of a module with a single light detector.

### 3.2. The role of light transport for the light signal

In terms of light production (Fig. 3), there is no significant difference between the modules. However, the light detection efficiency varies by nearly 50% (Fig. 4). The PMT measurements (which concern only the crystals) already show the variation; this suggests that the dominant part comes from the crystal itself absorbing the light.

In the common assumption, light absorbed in the crystal is converted to phonons and thus contributes to the phonon signal. Changing the energy balance in favor of the phonons would decrease the scintillation efficiency $\epsilon_s$ instead of—as we observed it—the detection efficiency $\epsilon_d$.

If instead the crystal re-absorbs light by exciting meta-stable states (whose relaxation time is longer than the phonon integration time of the measurements), the energy will not enter the phonon system on the time scale of a single event. Such a process only decreases the transport efficiency —corresponding to what we observed.

The cause for the different absorptivities might be the different treatments of the crystals. In case of *Wibke* and *Verena*, the TES was directly evaporated onto the crystal, while for the other detectors, a separate crystal carrying the TES was glued to the absorber crystal. Temperatures as high as needed for the evaporation process are known to deplete the crystal of oxygen, altering its optical properties [18].

### 3.3. The role of the light detectors for the light signal

With two light detectors combined, the double light detector module has a higher detection efficiency $\epsilon_d$ than any module with only a single light detector.

This indicates that the single light detectors are not capable of completely detecting the light in the setup presented. Hence it is possible to detect more light by e.g. increasing the size or absorptivity of the light detectors, or by changing the geometry of the module components, to favor propagation of light outside the crystal. We further explored [19] this possibility in the data-taking run following the one that provided data for this work.

## 4. Cross-check with a different method

### 4.1. Concept

In order to check the results for plausibility, we use the scintillating reflective housing of the detector modules as a reference light source. For all modules, the housing is made of the same scintillating reflector foil. Therefore, we consider it to emit equal amounts of light for a given energy deposited therein.

If individual modules detect different amounts of light from this reference light source, that must be due to different light detection efficiencies: the light can be not optimally reflected by the foil, not optimally transmitted by the crystal and/or not optimally absorbed by the light detector. Hence, we consider the amount of light detected from the reference light source $E_{L,\mathrm{ref}}$ to be proportional to the detection efficiency $\epsilon_d$ of the module.

As stated above, the amount of light which is detected due to a scintillation process in the crystal depends on the deposited energy, the scintillation efficiency of the crystal and the detection efficiency of the module. In order to fix all parameters but the scintillation efficiency, we divide the light signal $E_{L,\gamma}$ associated with a well-defined $\gamma$-line by the light signal of the reference light source $E_{L,\mathrm{ref}}$:

$$\rho = \frac{E_{L,\gamma}}{E_{L,\mathrm{ref}}} \qquad (10)$$

The light detectors might measure the light from the crystal with a different efficiency compared to light from the reference light source. Assuming that for all detector modules, this effect follows the same systematics, the fraction $\rho$ is proportional to the scintillation efficiency $\epsilon_s$.

### 4.2. Data

As in case of the main method, the data for the crystal scintillation were obtained from the $^{228}$Ac and $^{212}$Bi lines. A well-defined set of events which make the foil scintillate is caused by α-particles from the decay of $^{210}$Po on the surfaces of either the crystal or the housing (cf. [3]).[9]

### 4.3. Results

For identically designed modules (single light detector), the absolute values $\rho$ and $E_{L,\mathrm{ref}}$ from the cross-check method agree with the scintillation and detection efficiencies $\epsilon_s$ and $\epsilon_d$ from the main method. This behavior is visualized in Figs. 3 and 4, where the values for $\rho$ and $E_{L,\mathrm{ref}}$ (both times in red) were scaled with a constant factor determined by a least squares fit over all detector modules investigated.

## 5. Conclusion and outlook

The present work introduces a method for determining the energy distribution in a CRESST detector module. It uses in-situ data obtained at low temperatures, during a dark matter measurement campaign. We could show that the method can successfully distinguish between the efficiencies of light production and detection in a module. The results are reasonably consistent with a second method that provides a cross-check with different systematics.

We found that for γ-events, the crystals convert slightly less than 10% of the deposited energy into scintillation light. The variation between the individual detector modules is in the range of 10%. These results confirm previously published values.

In the modules we tested, only 20–30% of the produced light is detected, corresponding to a variation of nearly 50% among the modules. Measurements with the same crystals but without the module housings show the same variations. This suggests that characteristics of the crystals —instead of the module housings— dominate the detection efficiencies. A possible channel in which light disappears (explaining the low detection efficiencies) could be excitations of meta-stable states in the crystal.

The results obtained with the double light detector module indicate that the single light detector of a standard module does not absorb the entire amount of collectible light. Hence, apart from producing crystals that absorb less light, detecting more of the available light seems promising for further optimizing the detector modules. Our current [19] optimizations aim at both directions: via cuboid crystals (reducing the number of reflections within the crystal) and with larger-area light detectors.

## Acknowledgments

This research was supported by the German Federal Ministry of Science and Education (BMBF), the DFG cluster of excellence: Origin and Structure of the Universe, the DFG Transregio 27: Neutrinos and Beyond, the Helmholtz Alliance for Astroparticle Physics, the Maier-Leibnitz-Laboratorium (Garching) and the Science and Technology Facilities Council (STFC) UK.

## References

[1] G. Bertone, et al., Particle Dark Matter, Cambridge University Press, Cambridge, 2010.
[2] G. Angloher, A. Bento, C. Bucci, L. Canonica, A. Erb, F. von Feilitzsch, Iachellini N. Ferreiro, P. Gorla, A. Gütlein, D. Hauff, P. Huff, J. Jochum, M. Kiefer, C. Kister, H. Kluck, H. Kraus, J.-C. Lanfranchi, J. Loebell, A. Münster, F. Petricca, W. Potzel, F. Pröbst, F. Reindl, S. Roth, K. Rottler, C. Sailer, K. Schäffner, J. Schieck, J. Schmaler, S. Scholl, S. Schönert, W. Seidel, M. von Sivers, L. Stodolsky, C. Strandhagen, R. Strauss, A. Tanzke, M. Uffinger, A. Ulrich, I. Usherov, S. Wawoczny, M. Willers, M. Wüstrich, A. Zöller, Eur. Phys. J. C 74 (12) (2014) 3184, http://dx.doi.org/10.1140/epjc/s10052-014-3184-9 (arXiv:1407.3146).
[3] G. Angloher, M. Bauer, I. Bavykina, A. Bento, C. Bucci, C. Ciemniak, G. Deuter. F. von Feilitzsch, D. Hauff, P. Huff, C. Isaila, J. Jochum, M. Kiefer, M. Kimmerle, J.-C. Lanfranchi, F. Petricca, S. Pfister, W. Potzel, F. Pröbst, F. Reindl, S. Roth, K.Rottler, C. Sailer, K. Schäffner, J. Schmaler, S. Scholl, W. Seidel, M. von Sivers, L. Stodolsky, C. Strandhagen, R. Strau, A. Tanzke, I. Usherov, S. Wawoczny, M. Willers, A. Zöller, Eur. Phys. J. C 72 (4) (2012), http://dx.doi.org/10.1140/epjc/s10052-012-1971-8 (arXiv:1109.0702).
[4] F. Ferroni, Il Nuovo Cimento, 33 (5) (2010), http://dx.doi.org/10.1393/ncc/i2011-10696-1.
[5] H. Bhang, R.S. Boiko, D.M. Chernyak, J.H. Choi, S. Choi, F.A. Danevich, K.V. Efendiev, C. Enss, A. Fleischmann, A.M. Gangapshev, L. Gastaldo, A.M. Gezhaev, Y.S. Hwang, H. Jiang, W.G. Kang, V.V. Kazalov, N.D. Khanbekov, H.J. Kim, K.B. Kim, S.K. Kim, S.C. Kim, Y.D. Kim, Y.H. Kim, V.V. Kobychev, V.N. Kornoukhov, V. V. Kuzminov, V.M. Mokina, H.S. Lee, J.I. Lee, J.M. Lee, K.B. Lee, M.J. Lee, M.K. Lee, S.J. Lee, J. Li, X. Li, S.S. Myung, A.S. Nikolaiko, S. Olsen, S.I. Panasenko, H. Park, D. V. Poda, R.B. Podviyanuk, O.G. Polischuk, P.A. Polozov, S.S. Ratkevich, Y. Satou, J. H. So, K. Tanida, V.I. Tretyak, S.P. Yakimenko, Q. Yue, Y. Yuryev, J. Phys.: Conf. Ser. 375 (4) (2012) 042023 (URL ⟨http://stacks.iop.org/1742-6596/375/i=4/a=042023⟩).
[6] A. Barabash, D. Chernyak, F. Danevich, A. Giuliani, I. Ivanov, E. Makarov, M. Mancuso, S. Marnieros, S. Nasonov, C. Nones, E. Olivieri, G. Pessina, D. Poda, V. Shlegel, M. Tenconi, V. Tretyak, Y. Vasiliev, M. Velazquez, V. Zhdankov, Eur. Phys. J. C 74 (10) (2014), http://dx.doi.org/10.1140/epjc/s10052-014-3133-7 (arXiv:1405.6937. URL ⟨http://dx.doi.org/10.1140/epjc/s10052-014-3133-7⟩).
[7] F. Prö, M. Frank, S. Cooper, P. Colling, D. Dummer, P. Ferger, G. Forster, A. Nucciotti, W. Seidel, L. Stodolsky, J. Low Temp. Phys. 100 (1995) 69, http://dx.doi.org/10.1007/BF00753837.
[8] M.C. Gupta, R.D. MacFarlane, J. Inorg. Nucl. Chem. 32 (11) (1970) 3425, http://dx.doi.org/10.1016/0022-1902(70)80149-X.
[9] S. Chu, L. Ekström, R. Firestone, WWW Table of Radioactive Isotopes, vol. 2, 1999 (URL ⟨http://nucleardata.nuclear.lu.se/nucleardata/toi/⟩).
[10] V.B. Mikhailik, H. Kraus, Phys. Status Solidi B 247 (7) (2010) 1583, http://dx.doi.org/10.1002/pssb.200945500 (URL ⟨http://dx.doi.org/10.1002/pssb.200945500⟩).
[11] V. Tretyak, Astropart. Phys. 33 (2010) 40, http://dx.doi.org/10.1016/j.astropartphys.2009.11.002 (arXiv:0911.3041).
[12] J.B. Birks, The Theory and Practice of Scintillation Counting, Pergamon Press, Oxford, 1964.
[13] R. Strauss, G. Angloher, A. Bento, C. Bucci, L. Canonica, W. Carli, A. Erb, F. vonFeilitzsch, P. Gorla, A. Gütlein, D. Hauff, D. Hellgartner, J. Jochum, H. Kraus, J.-C. Lanfranchi, J. Loebell, A. Münster, F. Petricca, W. Potzel, F. Pröbst, F. Reindl, S. Roth, K. Rottler, C. Sailer, K. Schäffner, J. Schieck, S. Scholl, S. Schönert, W. Seidel, M. von Sivers, L. Stodolsky, C. Strandhagen, A. Tanzke, M. Uffinger, A. Ulrich, I. Usherov, S. Wawoczny, M. Willers, M. Wüstrich, A. Zöller, Eur. Phys. J. C 74 (7) (2014), http://dx.doi.org/10.1140/epjc/s10052-014-2957-5.
[14] T. Taulbee, B. Rooney, W. Mengesha, J. Valentine, IEEE Trans. Nucl. Sci. NS-44 (3) (1997) 489, http://dx.doi.org/10.1109/23.603696.
[15] M. Moszynski, Nucl. Instrum. Methods Phys. Res. Sect. A 505 (1–2) (2003) 101, http://dx.doi.org/10.1016/S0168-9002(03)01030-1 (Proceedings of the 10th Symposium on Radiation Measurements and Applications).
[16] R. Lang et al., Scintillator non-proportionality and gamma quenching in CaWO$_4$, München 2009 ⟨http://arxiv.org/abs/0910.4414⟩.
[17] C. Cozzini, G. Angloher, C. Bucci, F. von Feilitzsch, D. Hauff, S. Henry, T. Jagemann, J. Jochum, H. Kraus, B. Majorovits, V. Mikhailik, J. Ninkovic, F. Petricca, W. Potzel, F. Pröbst, Y. Ramachers, W. Rau, M. Razeti, W. Seidel, M. Stark, L. Stodolsky, A.J.B. Tolhurst, W. Westphal, H. Wulandari, Phys. Rev. C 70 (6) (2004) 064606, http://dx.doi.org/10.1103/PhysRevC.70.064606 (arXiv:nucl-ex/0408006).
[18] J. Ninkovic, Investigation of CaWO$_4$ crystals for simultaneous phonon-light detection in the CRESST dark matter search (Ph.D. thesis), TU München, 2005.
[19] F. Reindl, G. Angloher, A. Bento, C. Bucci, L. Canonica, A. Erb, A. Ertl, F. v. Feilitzsch, P. Gorla, D. Hauff, J. Jochum, C. Kister, H. Kraus, J.-C. Lanfranchi, J. Loebell, A. Münster, F. Petricca, W. Potzel, F. Pröbst, S. Roth, K. Rottler, C. Sailer, K. Schäffner, W. Seidel, M. v. Sivers, L. Stodolsky, C. Strandhagen, R. Strauβ, A. Tanzke, M. Uffinger, A. Ulrich, I. Usherov, M. Wüstrich, S. Wawoczny, M. Willers, A. Zöller, status update on the CRESST dark matter search, in: Astroparticle, Particle, Space Physics and Detectors for Physics Applications, vol. 8, 2014, p. 290–296, http://dx.doi.org/10.1142/9789814603164_0045.

---

[9] Having a relative light yield of $\mathcal{R}_{Pb} \approx 0.014\%$ [3], the recoiling Pb-nuclei contribute an amount of light that is negligible compared to the αs.